\documentclass[pra,showpacs,twocolumn]{revtex4}
\usepackage{graphicx}

\begin{document}

\title{Beating quantum limits in optomechanical sensor by cavity detuning}

\author{O. Arcizet}
\author{T. Briant}
\author{A. Heidmann}
\author{M. Pinard}

\affiliation{Laboratoire Kastler Brossel, Case 74, 4 place
Jussieu, F75252 Paris Cedex 05, France}
\thanks{Unit\'{e} mixte de recherche du Centre National de la Recherche
Scientifique, de l'Ecole Normale Sup\'{e}rieure et de
l'Universit\'{e} Pierre et Marie Curie}
\homepage{www.spectro.jussieu.fr/Mesure}

\date{January 30, 2006}

\begin{abstract}
We study the quantum limits in an optomechanical sensor based on a
detuned high-finesse cavity with a movable mirror. We show that
the radiation pressure exerted on the mirror by the light in the
detuned cavity induces a modification of the mirror dynamics and
makes the mirror motion sensitive to the signal. This leads to an
amplification of the signal by the mirror dynamics, and to an
improvement of the sensor sensitivity beyond the standard quantum
limit, up to an ultimate quantum limit only related to the
mechanical dissipation of the mirror. This improvement is somewhat
similar to the one predicted in detuned signal-recycled
gravitational-waves interferometers, and makes a high-finesse
cavity a model system to test these quantum effects.
\end{abstract}


\maketitle

\section{Introduction} \label{sec:Intro}

Quantum noise of light is known to induce fundamental limits in
very sensitive optical measurements. As an example, the future
generations of gravitational-wave interferometers
\cite{Bradaschia90,Abramovici92,Fritschel02} will most probably be
confronted to quantum effects of radiation pressure. A
gravitational wave induces a differential variation of the optical
paths in the two arms of a Michelson interferometer. The detection
of the phase difference between the two paths is ultimately
limited by two quantum noise sources: the phase fluctuations of
the incident laser beam and the radiation pressure effects which
induce unwanted mirror displacements in the interferometer. A
compromise between these noises leads to the so-called standard
quantum limit for the sensitivity of the measurement
\cite{Caves81,Jaekel90,Braginsky92}.

Number of quantum noise reduction schemes have been proposed which
rely on the injection of squeezed states of light in the
interferometer \cite{Xiao87,Grangier87,McKenzie02}, or on the
quantum correlations induced by radiation pressure between phase and
intensity fluctuations in the interferometer \cite{Kimble02}. The
possibility to implement these techniques in real interferometers
gave rise to new methods such as the quantum locking of mirrors
\cite{Courty03} or the detuning of the signal recycling cavity
\cite{Buonanno01,Harms03}.

It seems important to find simple systems where similar quantum
effects can be produced and characterized in order to test these
effects in tabletop experiments. From this point of view,
high-finesse optical cavities with movable mirrors have
interesting potentialities since they exhibit similar quantum
limits. Several schemes involving such cavities have been proposed
either to create non-classical states of both the radiation field
\cite{Fabre94,Mancini94} and of the mirror motion
\cite{Bose99,Mancini02,Pinard05}, or to perform quantum
nondemolition measurements \cite{Heidmann97}. Recent progress in
low-noise laser sources and low-loss mirrors have made the field
experimentally accessible \cite{Hadjar99,Cohadon99,Tittonen99}.

We study in this paper the quantum effects in a detuned cavity and
the possibility to beat the standard quantum limit. As for
signal-recycled interferometers \cite{Buonanno02}, the detuning of
the cavity induces a modification of the mechanical dynamics of
the mirror, known as optical spring. This effect may improve the
sensitivity beyond the standard quantum limit since it changes the
mechanical rigidity of the mirror without any additional noise
\cite{Braginsky99,Braginsky01}. The optical spring has already
been observed in a Fabry-Perot cavity \cite{Sheard04}, and studied
both theoretically \cite{Braginsky02} and experimentally
\cite{Schediwy04} for its role in parametric instabilities.

We perform a full quantum treatment of a detuned cavity with a
movable mirror. We show that the sensitivity of the measurement of a
cavity length variation can be made better than the standard quantum
limit. From a careful analysis of the mirror dynamics, we find that
it is not only attributed to the optical spring, but also to the
fact that the mirror becomes sensitive to the signal through the
radiation pressure exerted on the mirror. We show that the mirror
motion can amplify the signal, thus increasing the sensitivity up to
an ultimate quantum limit only related to the dissipation mechanisms
of the mechanical motion \cite{Jaekel90}. We finally study the
influence of a finite cavity bandwidth and obtain dual sensitivity
peaks similar to the ones obtained for detuned signal-recycled
interferometers \cite{Buonanno01}.

\section{Optomechanical coupling in a detuned cavity}
\label{sec:COM}

\begin{figure}
\includegraphics[width=8cm]{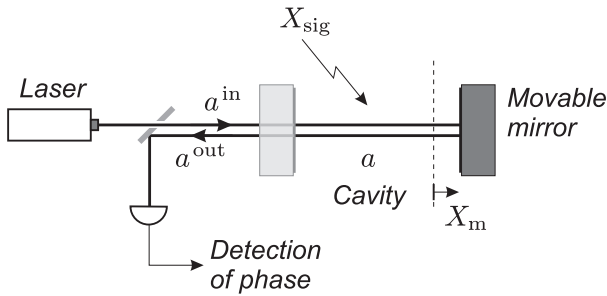}
\caption{A length variation $X_{\rm sig}$ is measured by a
single-ended Fabry-Perot cavity through the phase shift induced on
the reflected field $a^{\rm out}$. Radiation pressure effects are
taken into account via the displacement $X_{\rm m}$ of the movable
end mirror.} \label{fig:cavity}
\end{figure}

We consider the single-port cavity shown in Fig. \ref{fig:cavity}
with a partially transmitting front mirror and a totally reflecting
end mirror. A probe laser beam is sent in the cavity and the phase
of the reflected field is monitored by a homodyne detection. We
consider in the following the motion of a single mirror assuming the
front mirror is fixed, and we note $X_{\rm m}$ the displacement of
the movable end mirror. We study the response of the system to a
signal described as a variation $X_{\rm sig}$ of the cavity length.
It can either be a physical variation of the cavity length due for
example to an external force applied on the movable mirror, or an
apparent variation such as the one produced by a gravitational wave
in a gravitational-wave interferometer. The time-dependent cavity
length $L$ is then given by
\begin{equation}
L(t)=L_0+X_{\rm m}\left(t\right)+X_{\rm sig}\left(t\right),
\label{eq:L}
\end{equation}
where $L_0$ is the cavity length without signal and for a mirror
at rest.

For a nearly resonant high-finesse cavity, the intracavity field
mode described by the annihilation and creation operators
$a\left(t\right)$ and $a^{\dagger}\left(t\right)$ is related to the
input and output fields $a^{\rm in}\left(t\right)$ and $a^{\rm
out}\left(t\right)$ by
\begin{eqnarray}
\tau\frac{da\left(t\right)}{dt}&=&-\left[\gamma-i\psi\left(t\right)\right]a\left(t\right)+
\sqrt{2\gamma}a^{\rm in}\left(t\right), \label{eq:InOut1}\\
a^{\rm out}\left(t\right)&=&-a^{\rm in}\left(t\right)+
\sqrt{2\gamma}a\left(t\right), \label{eq:InOut2}
\end{eqnarray}
where $\gamma$ is the damping rate of the cavity assumed to be
small compared to 1, $\tau$ is the cavity round trip time, and
$\psi\left(t\right)$ is the time-dependent detuning of the cavity
related to the cavity length by
\begin{equation}
\psi\left(t\right) \equiv 2k L\left(t\right) \left[2\pi\right],
\label{eq:psi}
\end{equation}
where $k$ is the field wavevector.

The intracavity field induces a radiation pressure force $F_{\rm
rad}$ on the mirror which is proportional to the field intensity,
\begin{equation}
F_{\rm rad}\left(t\right) = 2\hbar k I\left(t\right),
\label{eq:Frad}
\end{equation}
where the intracavity intensity $I=\left|a\right|^2$ is normalized
as a photon flux. In the framework of linear response theory
\cite{Landau58}, the Fourier transform $X_{\rm
m}\left[\Omega\right]$ of the mirror displacement at frequency
$\Omega$ linearly depends on the applied force
$F\left[\Omega\right]$,
\begin{equation}
X_{\rm m}\left[\Omega\right] = \chi\left[\Omega\right]
F\left[\Omega\right], \label{eq:Xm}
\end{equation}
where $\chi\left[\Omega\right]$ is the mechanical susceptibility
of the mirror. Assuming that the mirror motion can be described as
the one of a single harmonic oscillator with a resonance frequency
$\Omega_{\rm M}$, a mass $M$, and a damping rate $\Gamma$, the
susceptibility has the simple form
\begin{equation}
\chi\left[\Omega\right] = \frac1{M\left(\Omega_{\rm M}^2 -
\Omega^2 - i\Gamma\Omega\right)}. \label{eq:chi}
\end{equation}

The steady state is obtained by cancelling the time derivative in
(\ref{eq:InOut1}). One gets the steady states $\overline a$ and
$\overline a^{\rm out}$ of the intracavity and output fields as a
function of the incident mean field $\overline a^{\rm in}$ and the
mean detuning $\overline \psi$ of the cavity,
\begin{equation}
\overline a = \frac {\sqrt{2\gamma}} {\gamma-i\overline \psi}
\overline a^{\rm in} = \frac {\sqrt{2\gamma}} {\gamma+i\overline
\psi} \overline a^{\rm out}. \label{eq:a_mean}
\end{equation}
As expected for a lossless cavity, the outgoing mean intensity
$\left|\overline a^{\rm out}\right|^2$ is equal to the incident
one $\left|\overline a^{\rm in}\right|^2$. For a non-zero detuning
$\overline \psi$, the mean fields  $\overline a^{\rm in}$,
$\overline a$, and $\overline a^{\rm out}$ have different phases.
We choose by convention the arbitrary global phase of the fields
in such a way that the intracavity field $\overline a$ is real.
The phases $\theta^{\rm in}$ and $\theta^{\rm out}$ of the input
and output mean fields are then given by
\begin{equation}
e^{-i\theta^{\rm in}}=\frac{\gamma-i\overline \psi}
{\sqrt{\gamma^2+\overline \psi^2}}\qquad,\qquad e^{-i\theta^{\rm
out}}=\frac{\gamma+i\overline \psi} {\sqrt{\gamma^2+\overline
\psi^2}}. \label{eq:theta}
\end{equation}

According to eqs. (\ref{eq:L}) to (\ref{eq:Xm}), the mean detuning
$\overline \psi$ depends on the intracavity intensity through the
mirror recoil induced by the intracavity radiation pressure,
\begin{equation}
\overline \psi = \psi_0 + \hbar \kappa^2 \chi[0],
\label{eq:psi_mean}
\end{equation}
where $\psi_0 \equiv 2kL_0 \left[2\pi\right]$ is the detuning
without light and $\kappa = 2k \left|\overline a\right|$. The
coupled equations (\ref{eq:a_mean}) and (\ref{eq:psi_mean}) give a
third order relation between $\overline a$ and $\overline \psi$
which leads to the well known bistable behavior of a cavity with a
movable mirror \cite{Dorsel83}. The stability condition of the
system can be written as
\begin{equation}
\gamma^2+\overline \psi^2+2\hbar \kappa^2 \chi \left[0\right]
\overline \psi > 0, \label{eq:stability1}
\end{equation}

\section{Mirror dynamics} \label{sec:Mirror}

We derive in this section the basic input-output relations for the
fluctuations and we study the modification of the mirror dynamics
induced by the radiation pressure in the cavity. We will show that
the mechanical response of the mirror to an external force is
modified by the optomechanical coupling with the light. It can be
described by an effective mechanical susceptibility similar to the
one obtained with an active control of the mirror by a feedback loop
\cite{Cohadon99,Pinard00,Courty01}.

The linearization of the Fourier transform of eq.
(\ref{eq:InOut1}) around the mean state gives the intracavity
field $a\left[\Omega\right]$, at a given frequency $\Omega$, as a
function of the incident field fluctuations $a^{\rm in}$ and the
cavity length variations $X_{\rm m}$ and $X_{\rm sig}$,
\begin{eqnarray}
\left(\gamma-i\overline \psi-i\Omega \tau\right)
a\left[\Omega\right] &=& \sqrt{2\gamma} a^{\rm
in}\left[\Omega\right] + i\kappa X_{\rm m}\left[\Omega\right]
\nonumber \\
&&+ i\kappa X_{\rm sig}\left[\Omega\right]. \label{eq:InOut3}
\end{eqnarray}
According to eq. (\ref{eq:Frad}), the radiation pressure $F_{\rm
rad}\left[\Omega\right]$ depends on the intensity fluctuations of
the intracavity field at frequency $\Omega$. From eq.
(\ref{eq:InOut3}) it can be written as the sum of three forces,
\begin{eqnarray}
F_{\rm rad}^{\left({\rm in}\right)}\left[\Omega\right] &=& \hbar
\kappa \sqrt{\frac{2\gamma}{\gamma^2+\overline \psi^2}} \left(
\frac{\gamma^2+\overline \psi^2-i\gamma\Omega\tau}{\Delta}
p^{\rm in}\left[\Omega\right] \right. \nonumber \\
&&-\left. \frac{i\overline \psi \Omega\tau}{\Delta} q^{\rm
in}\left[\Omega\right]\right), \label{eq:Frad_in} \\
F_{\rm rad}^{\left({\rm m}\right)}\left[\Omega\right] &=& -2\hbar
\kappa^2 \frac{\overline \psi}{\Delta} X_{\rm
m}\left[\Omega\right], \label{eq:Frad_m} \\
F_{\rm rad}^{\left({\rm sig}\right)}\left[\Omega\right] &=&
-2\hbar \kappa^2 \frac{\overline \psi}{\Delta} X_{\rm
sig}\left[\Omega\right], \label{eq:Frad_sig}
\end{eqnarray}
where $\Delta = \left(\gamma-i\Omega\tau\right)^2+\overline
\psi^2$ and the operators $p^{\rm in}\left[\Omega\right]$ and
$q^{\rm in}\left[\Omega\right]$ correspond to the amplitude and
phase quadratures of the incident field, respectively,
\begin{eqnarray}
p^{\rm in}\left[\Omega\right] &=& e^{i\theta^{\rm in}} a^{\rm
in}\left[\Omega\right] + e^{-i\theta^{\rm in}}{a^{\rm
in}}^{\dagger}\left[\Omega\right], \label{eq:quad_p} \\
q^{\rm in}\left[\Omega\right] &=& -ie^{i\theta^{\rm in}} a^{\rm
in}\left[\Omega\right] + ie^{-i\theta^{\rm in}}{a^{\rm
in}}^{\dagger}\left[\Omega\right], \label{eq:quad_q}
\end{eqnarray}
(same definitions hold for the intracavity quadratures with an angle
$\theta=0$ and for the reflected ones with the angle $\theta^{\rm
out}$). The first force $F_{\rm rad}^{\left({\rm in}\right)}$
represents the radiation pressure induced by the quantum
fluctuations of the incident field. It is the usual force obtained
in the case of a resonant cavity, which is responsible for the
generation of squeezing in a cavity with a movable mirror
\cite{Fabre94}. Since it induces a displacement of the mirror
proportional to the field fluctuations, it is also responsible for
the standard quantum limits in interferometric measurements
\cite{Caves81,Jaekel90,Braginsky92}. For a resonant cavity
($\overline \psi=0$), this force only depends on the incident
intensity fluctuations $p^{\rm in}$ filtered  by the cavity
bandwidth $\Omega_{\rm cav}=\gamma /\tau$.

The two other forces $F_{\rm rad}^{\left({\rm m}\right)}$ and
$F_{\rm rad}^{\left({\rm sig}\right)}$ only exist when the cavity
is detuned ($\overline \psi\neq 0$). In that case the working
point of the cavity is on one side of the Airy peak. According to
eqs. (\ref{eq:psi}) and (\ref{eq:a_mean}), the intracavity
intensity depends on the cavity length variations with a slope
\begin{equation}
\frac{d\overline I}{dX}=-2\kappa \frac{\overline
\psi}{\gamma^2+\overline \psi^2}\overline a. \label{eq:AirySlope}
\end{equation}
Any length variation changes the intracavity intensity and induces
a variation of the radiation pressure exerted on the mirror. This
variation corresponds to the forces $F_{\rm rad}^{\left({\rm
m}\right)}$ and $F_{\rm rad}^{\left({\rm sig}\right)}$ which are
actually proportional to the slope (\ref{eq:AirySlope}) of the
Airy peak,
\begin{equation}
F_{\rm rad}^{\left(j\right)}\left[\Omega \right] = 2 \hbar k
\frac{\gamma^2+\overline \psi^2}{\Delta} \frac{d\overline I}{dX}
X_j\left[\Omega\right], \label{eq:FradSlope}
\end{equation}
where $j=\left({\rm m},{\rm sig}\right)$. The first fraction in
(\ref{eq:FradSlope}) is a low-pass filter associated with the
cavity storage time. The sign of the forces depends on the sign of
the slope (\ref{eq:AirySlope}). Depending on the sign of
$\overline \psi$, the force $F_{\rm rad}^{\left({\rm m}\right)}$
is either a repulsive or an attractive force, and the force
$F_{\rm rad}^{\left({\rm sig}\right)}$ induces a mirror
displacement which may either amplify or compensate the signal
$X_{\rm sig}$. We will see in the next section that this signal
amplification by the mirror motion is at the basis of the
sensitivity improvement obtained with a detuned cavity.

The force $F_{\rm rad}^{\left({\rm m}\right)}$ is proportional to
the mirror displacement $X_{\rm m}$. Its effect is to change the
mechanical response of the mirror to an external force which is
now given by eq. (\ref{eq:Xm}) with an effective susceptibility
$\chi_{\rm eff}$ related to the free susceptibility $\chi$ by
\begin{equation}
\chi_{\rm eff}^{-1}\left[\Omega\right] =
\chi^{-1}\left[\Omega\right] + 2\hbar \kappa^2 \frac{\overline
\psi}{\Delta}. \label{eq:chieff}
\end{equation}
If the frequencies $\Omega$ and $\Omega_{\rm M}$ are much smaller
than the cavity bandwidth $\Omega_{\rm cav}$, the additional term in
(\ref{eq:chieff}) is real. As a consequence, its effect is to change
the spring constant of the mechanical motion \cite{Braginsky99},
that is to shift the resonance frequency $\Omega_{\rm M}$ of the
oscillator [eq. (\ref{eq:chi})], either to low or high frequencies
depending on the sign of $\overline \psi$. If the frequencies
$\Omega$ and $\Omega_{\rm M}$ are of the order of $\Omega_{\rm
cav}$, the additional term in (\ref{eq:chieff}) becomes complex and
also changes the imaginary part of the susceptibility. If we
consider a mirror with a high quality factor ($\Gamma \ll
\Omega_{\rm M}$), the mechanical response (\ref{eq:chieff}) can
still be considered as Lorentzian with an effective damping
$\Gamma_{\rm eff}$ given by
\begin{equation}
\Gamma_{\rm eff} = \Gamma - \frac{4\hbar \kappa^2}{M\Omega_{\rm
cav}} \frac{\gamma^2 \overline \psi}{\left|\Delta\right|^2},
\label{eq:Gammaeff}
\end{equation}
where the denominator $\Delta$ is estimated at frequency
$\Omega_{\rm M}$. The mechanical resonance is widened or narrowed
depending on the sign of $\overline \psi$.

The coupling with the intracavity field thus changes the dynamics of
the mirror, both via its spring constant and its damping. The effect
is somewhat similar to the one obtained with an external feedback
control. In both cases it is for example possible to carry out a
cold damping mechanism which increases the damping without adding
extra thermal fluctuations, thus leading to a reduction of the
effective temperature of the mirror \cite{Cohadon99,Metzger04}. In
this paper we will use these effects in another way, in order to
amplify the response of the mirror to an external force. Together
with the sensitivity of the mirror motion to the signal via the
force $F_{\rm rad}^{\left({\rm sig}\right)}$, this will allow us to
greatly amplify the sensitivity of the cavity to the signal.

Let us note that the modification of the dynamics can lead to an
instability where the mirror enters a self-oscillating regime. The
dynamic stability condition is usually given by the Ruth-Hurwitz
criterium applied to the determinant of the linear relations between
the field and the mirror position \cite{Haken70,Fabre94}. It is
actually equivalent to the condition that the mirror motion has to
be characterized by a positive damping in order to have a
non-divergent motion,
\begin{equation}
\Gamma_{\rm eff} > 0. \label{eq:Stability2}
\end{equation}

\section{Sensitivity of the measurement} \label{sec:Sensitivity}

We now determine the sensitivity of the measurement and how it is
modified by the cavity detuning. We consider for simplicity that
the low-pass filtering by the cavity can be neglected, that is all
frequencies of interest ($\Omega$, $\Omega_{\rm M}$) are much
smaller than $\Omega_{\rm cav}$. This assumption will be relaxed
in section \ref{sec:FiniteBandwidth}.

The measurement is done by monitoring the phase quadrature $q^{\rm
out}$ of the field reflected by the cavity, as shown in Fig.
\ref{fig:cavity}. The linearized input-output relations for the
field are deduced from eqs. (\ref{eq:InOut2}) and
(\ref{eq:InOut3}),
\begin{eqnarray}
p^{\rm out}\left[\Omega\right] &=& p^{\rm in}\left[\Omega\right],
\label{eq:pout} \\
q^{\rm out}\left[\Omega\right] &=& q^{\rm in}\left[\Omega\right] +
2\xi \left(X_{\rm m}\left[\Omega\right]+X_{\rm
sig}\left[\Omega\right]\right), \label{eq:qout}
\end{eqnarray}
where the optomechanical coupling parameter $\xi$ is given by
\begin{equation}
\xi = \sqrt{\frac{2\gamma}{\gamma^2+\overline\psi^2}}\kappa = 2k
\frac{2\gamma}{\gamma^2+\overline\psi^2} \left|\overline a^{\rm
in}\right|. \label{eq:xi}
\end{equation}
The working point of the cavity will be defined in the following
by the two independent parameters $\overline \psi$ and $\xi$.
Other parameters such as the incident and intracavity intensities
can be deduced from eqs. (\ref{eq:a_mean}) and (\ref{eq:xi}).

Equations (\ref{eq:pout}) and (\ref{eq:qout}) show that as long as
we consider the quasi-static regime $\Omega\ll\Omega_{\rm cav}$,
the input-output relations are similar for a resonant and a
detuned cavity. Due to the preservation of the photon flux in a
lossless single-ended cavity, the reflected amplitude fluctuations
are equal to the incident ones and only the reflected phase
quadrature is sensitive to the variation $X_{\rm m}+X_{\rm sig}$
of the cavity length. This variation is superimposed to the
incident phase noise $q^{\rm in}$.

The mirror is submitted to the radiation pressure of the
intracavity field. As shown in the previous section, the response
to the forces $F_{\rm rad}^{\left({\rm in}\right)}$ and $F_{\rm
rad}^{\left({\rm sig}\right)}$ [eqs. (\ref{eq:Frad_in}) and
(\ref{eq:Frad_sig})] is characterized by the effective mechanical
susceptibility $\chi_{\rm eff}$ [eq. (\ref{eq:chieff})]. In the
limit $\Omega \ll \Omega_{\rm cav}$, this susceptibility and the
resulting motion are given by
\begin{eqnarray}
\chi_{\rm eff}^{-1}\left[\Omega\right] &=&
\chi^{-1}\left[\Omega\right] + \hbar \xi^2 \frac{\overline
\psi}{\gamma}, \label{eq:chieff2} \\
X_{\rm m}\left[\Omega\right] &=& \chi_{\rm eff}\left[\Omega\right]
\left(\hbar \xi p^{\rm in}\left[\Omega\right] - \hbar \xi^2
\frac{\overline \psi}{\gamma} X_{\rm
sig}\left[\Omega\right]\right). \label{eq:Xm2}
\end{eqnarray}
The mirror motion reproduces the signal $X_{\rm sig}$ with a
dynamics characterized by the effective susceptibility $\chi_{\rm
eff}$. Depending on the sign of $\overline \psi \chi_{\rm
eff}\left[\Omega\right]$, the mirror displacement is in phase or
out of phase with the signal, thus leading to an amplification or
a reduction of the signal in the output phase quadrature. This
quadrature is obtained from eqs. (\ref{eq:qout}) and
(\ref{eq:Xm2}),
\begin{eqnarray}
q^{\rm out}\left[\Omega\right] &=& q^{\rm in}\left[\Omega\right] +
2\hbar \xi^2 \chi_{\rm eff}\left[\Omega\right] p^{\rm
in}\left[\Omega\right] \nonumber \\
&&+ 2\xi \frac{\chi_{\rm eff}\left[\Omega\right]}
{\chi\left[\Omega\right]} X_{\rm sig}\left[\Omega\right].
\label{eq:qout2}
\end{eqnarray}
The signal $X_{\rm sig}$ is amplified by the coupling parameter
$\xi$ and by the dynamics of the mirror $\chi_{\rm eff}/\chi$
[last term in (\ref{eq:qout2})]. The signal is superimposed to two
noises respectively proportional to the phase and amplitude
incident fluctuations [first terms in (\ref{eq:qout2})]. These
noises are nothing but the usual phase noise and radiation
pressure noise in interferometric measurements.

It is instructive to compare the cases of detuned and resonant
cavities. For a resonant cavity, there is no modification of the
mechanical susceptibility ($\chi_{\rm eff} = \chi$) and the mirror
motion does not depend on the signal. The output quadrature is
simply obtained from eq. (\ref{eq:qout2}) by replacing $\chi_{\rm
eff}$ by $\chi$. There is no amplification of the signal and the
minimum noise corresponds to the standard quantum limit which is
reached when both the phase and radiation pressure noises are of the
same order, that is $\hbar \xi^2 \left|\chi\right| \approx 1$.

For a detuned cavity, the signal is amplified by the ratio
$\left|\chi_{\rm eff}/\chi\right|$, and the radiation pressure noise
is also increased by the same factor [second term in eq.
(\ref{eq:qout2})]. As long as we are only concerned by the noises,
the system is thus equivalent to a resonant cavity with a mirror
having an effective susceptibility $\chi_{\rm eff}$. Due to the
signal amplification, this is no longer true if we are looking at
the signal to noise ratio. The sensitivity of the measurement can be
increased beyond the standard quantum limit by choosing the
optomechanical parameters $\xi$ and $\overline \psi$ in such a way
that the signal is amplified ($\left|\chi_{\rm eff}\right| >
\left|\chi\right|$), whereas the quantum noises are still at the
standard quantum limit, that is $\hbar \xi^2 \left|\chi_{\rm
eff}\right| \approx 1$.

To derive a more precise evaluation of the sensitivity
improvement, we define an estimator $\widehat X_{\rm sig}$ of the
signal, equal to the measured quadrature $q^{\rm out}$ normalized
as the length variation $X_{\rm sig}$,
\begin{equation}
\widehat X_{\rm sig}\left[\Omega\right] = \frac1{2\xi}
\frac{\chi\left[\Omega\right]}{\chi_{\rm eff}\left[\Omega\right]}
q^{\rm out}\left[\Omega\right]. \label{eq:estimator}
\end{equation}
From eq. (\ref{eq:qout2}), this estimator appears as the sum of
the signal $X_{\rm sig}$ and two equivalent input noises
proportional to the incident fluctuations $q^{\rm in}$ and $p^{\rm
in}$. The sensitivity of the measurement is limited by the
spectrum $S_{\rm sig}\left[\Omega\right]$ of these noises. For a
coherent incident beam, the quantum fluctuations of the two
incident quadratures $p^{\rm in}$ and $q^{\rm in}$ are two
independent white noises with a unity spectrum ($S_p^{\rm
in}\left[\Omega\right]=S_q^{\rm in}\left[\Omega\right]=1$). The
equivalent noise spectrum $S_{\rm sig}\left[\Omega\right]$ is then
given by
\begin{equation}
S_{\rm sig}\left[\Omega\right] = \hbar
\left|\chi\left[\Omega\right]\right|
\left|\frac{\chi\left[\Omega\right]} {\chi_{\rm
eff}\left[\Omega\right]}\right|
\frac{\zeta\left[\Omega\right]+1/\zeta\left[\Omega\right]}{2},
\label{eq:Ssig}
\end{equation}
where the dimensionless parameter $\zeta$ is defined as
\begin{equation}
\zeta\left[\Omega\right] = 2\hbar\xi^2\left|\chi_{\rm
eff}\left[\Omega\right]\right|. \label{eq:zeta}
\end{equation}
The last fraction in eq. (\ref{eq:Ssig}) is always greater than 1
and reaches its minimum for $\zeta\left[\Omega\right] = 1$. In
that case, the phase and radiation pressure noises are equal and
their sum is minimum. For a resonant cavity, this corresponds to
the standard quantum limit which is reached at a given frequency
$\Omega$ for the following value $\xi_{\rm SQL}
\left[\Omega\right]$ of the optomechanical parameter, and
corresponds to a minimum noise level $S_{\rm sig}^{\rm
SQL}\left[\Omega\right]$ given by
\begin{eqnarray}
\xi_{\rm SQL}\left[\Omega\right] &=& \frac1{\sqrt{2 \hbar
\left|\chi\left[\Omega\right]\right|}}, \label{eq:xiSQL} \\
S_{\rm sig}^{\rm SQL}\left[\Omega\right] &=& \hbar
\left|\chi\left[\Omega\right]\right|. \label{eq:SsigSQL}
\end{eqnarray}

It is clear from eq. (\ref{eq:Ssig}) that the standard quantum limit
is not a fundamental limit. It is possible to go beyond this limit
with a detuned cavity, by choosing the optomechanical parameters so
that $\zeta\left[\Omega\right]\simeq 1$ and $\left|\chi_{\rm
eff}\left[\Omega\right]\right| >
\left|\chi\left[\Omega\right]\right|$. The sensitivity is then
increased by the amplification factor $\left|\chi_{\rm
eff}\left[\Omega\right] / \chi\left[\Omega\right]\right|$ given by
[see eq. (\ref{eq:chieff2})],
\begin{equation}
\left|\chi_{\rm eff}\left[\Omega\right] / \chi\left[\Omega\right]
\right| = \left|1 + \hbar \xi^2 \frac{\overline \psi}{\gamma}
\chi\left[\Omega\right] \right|^{-1}. \label{eq:amplification}
\end{equation}
Note that the term inside the absolute value, taken at frequency
$\Omega=0$, exactly corresponds to the stability condition of the
bistable behavior [eq. (\ref{eq:stability1})]. This term is
strictly positive in the stable domain, thus preventing the
amplification factor to diverge. We study in the next sections the
sensitivity improvement in two particular cases of experimental
interest, corresponding to a frequency $\Omega$ either below or
beyond the mechanical resonance frequency $\Omega_{\rm M}$.

\section{Sensitivity improvement at low frequency}
\label{sec:LowFreq}

We first consider the sensitivity improvement at frequency lower
than the mechanical resonance frequency. This situation is of
interest for the displacement measurements made with small and
compact high-finesse cavities, where the radiation pressure effects
are mainly due to the excitation of high-frequency internal acoustic
modes of the mirrors \cite{Pinard00,Briant03}. The susceptibility
$\chi\left[\Omega\right]$ at frequency well below the mechanical
resonance can be approximated as a real and positive expression [eq.
(\ref{eq:chi})],
\begin{equation}
\chi\left[\Omega \ll \Omega_{\rm M}\right] \simeq
\chi\left[0\right] = \frac1{M\Omega_{\rm M}^2}. \label{eq:chilow}
\end{equation}
According to eq. (\ref{eq:amplification}), $\overline \psi$ has
then to be negative in order to obtain an amplification factor
$\left|\chi_{\rm eff} / \chi\right|$ larger than 1. For any
arbitrary negative value of the detuning $\overline \psi$, the
condition $\zeta\left[0\right] = 1$ is reached for the value of
the optomechanical parameter given by
\begin{equation}
\xi^2 = \frac{\xi_{\rm SQL}^2\left[0\right]}{1-\overline \psi /
2\gamma}, \label{eq:xilow}
\end{equation}
and the corresponding noise spectrum is equal to
\begin{equation}
\frac{S_{\rm sig}\left[0\right]}{S_{\rm sig}^{\rm SQL}
\left[0\right]} = \frac1{\left|\chi_{\rm eff}\left[0\right] /
\chi\left[0\right]\right|} = \frac1{1-\overline \psi / 2\gamma}.
\label{eq:Ssiglow}
\end{equation}
It is then possible to arbitrarily reduce the equivalent input
noise and to increase the sensitivity by choosing a large negative
detuning. Note that although the optomechanical parameter $\xi$
given by eq. (\ref{eq:xilow}) decreases as the noise spectrum, it
corresponds to a larger incident intensity [see eq.
(\ref{eq:xi})]. Increasing the sensitivity thus requires a larger
input power.

\begin{figure}
\includegraphics[width=8cm]{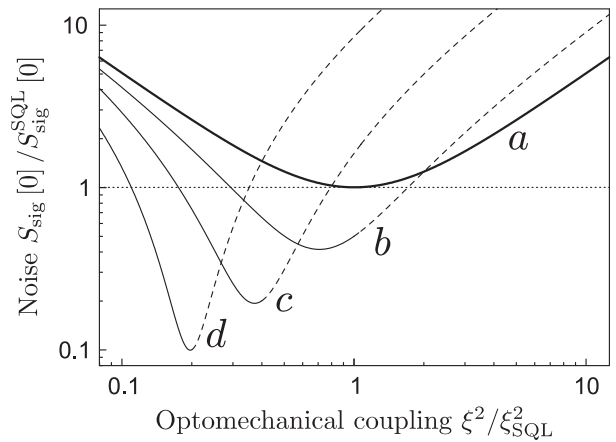}
\caption{Equivalent input noise $S_{\rm sig}$ at low frequency as
a function of the optomechanical parameter $\xi^2$, normalized to
the SQL values $\xi_{\rm SQL}^2$ and $S_{\rm sig}^{\rm SQL}$.
Curves {\it a} to {\it d} are plotted for a normalized detuning
$\overline \psi/\gamma$ equal to 0, $-2$, $-5$, and $-10$,
respectively. Dashed lines correspond to the unstable domain.}
\label{fig:lowfreq}
\end{figure}

Figure \ref{fig:lowfreq} shows the equivalent input noise $S_{\rm
sig}\left[0\right]$ as a function of the optomechanical parameter
$\xi$, for different values of the detuning $\overline \psi$. Curve
{\it a} is obtained at resonance ($\overline \psi = 0$). The noise
reaches the standard quantum limit for $\xi = \xi_{\rm SQL}$ and is
larger than this limit elsewhere. Since $\xi^2$ is proportional to
the incident intensity [eq. (\ref{eq:xi})], the additional noise for
$\xi < \xi_{\rm SQL}$ corresponds to the phase noise which is
dominant at low intensity, whereas the additional noise for $\xi >
\xi_{\rm SQL}$ corresponds to the radiation pressure noise, dominant
at high intensity. The behavior is similar for a detuned cavity
(curves {\it b} to {\it d}), with a minimum noise reached at
decreasing values of $\xi$ as the detuning increases. The minimum
noise is actually better than the one given by eq.
(\ref{eq:Ssiglow}). A more accurate optimization of the noise
spectrum (\ref{eq:Ssig}) leads to
\begin{eqnarray}
\xi_{\rm min}^2 &=& \frac{\xi_{\rm SQL}^2\left[0\right]}
{\sqrt{1+\left(\overline \psi/2\gamma\right)^2}},
\label{eq:xilow2} \\
\frac{S_{\rm sig}^{\rm min}\left[0\right]}{S_{\rm sig}^{\rm SQL}
\left[0\right]} &=& \sqrt{1+\left(\overline \psi /
2\gamma\right)^2} + \overline \psi / 2\gamma, \label{eq:Ssiglow2}
\end{eqnarray}
which tends to $\gamma/\left|\overline \psi\right|$ for large
detunings. As an example, the curve {\it d} corresponding to a
detuning $\overline \psi = -10\gamma$ exhibits a noise reduction by
a factor 10. Finally note that as the amplification by the mirror
increases with the detuning, the optimum working point becomes
nearer and nearer to the unstable domain shown as dashed curves in
Fig. \ref{fig:lowfreq}. It however always stays in the stable domain
of the bistable behavior given by eq. (\ref{eq:stability1}).

\section{Sensitivity improvement at high frequency}
\label{sec:HighFreq}

\begin{figure}
\includegraphics[width=8cm]{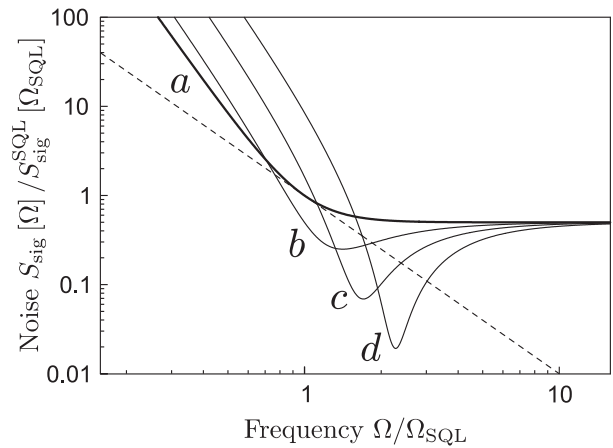}
\caption{Equivalent input noise $S_{\rm sig}\left[\Omega\right]$
at high frequency as a function of frequency $\Omega$, normalized
to the SQL values $\Omega_{\rm SQL}$ and $S_{\rm sig}^{\rm
SQL}\left[\Omega_{\rm SQL}\right]$. Curves {\it a} to {\it d} are
plotted for the same optomechanical parameter $\xi$ and for
increasing normalized detunings $\overline \psi/\gamma$, equal to
0, $2$, $5$, and $10$, respectively. Dashed line corresponds to
the standard quantum limit.} \label{fig:highfreq}
\end{figure}

We now study the sensitivity improvement at frequency larger than
the mechanical resonance frequency. This situation corresponds for
example to gravitational-wave interferometers where the main motion
of the mirror is due to the pendular suspension which has very low
resonance frequencies \cite{Bradaschia90,Abramovici92}. In this
case, the susceptibility $\chi\left[\Omega\right]$ can be
approximated as a real but negative expression [eq. (\ref{eq:chi})],
\begin{equation}
\chi\left[\Omega \gg \Omega_{\rm M}\right] \simeq
-\frac1{M\Omega^2}, \label{eq:chihi}
\end{equation}
so that the amplification factor $\left|\chi_{\rm eff} /
\chi\right|$ is now larger than 1 for a positive detuning
$\overline \psi$. Since the susceptibility is frequency dependent,
the condition $\zeta\left[\Omega\right] = 1$ can be satisfied at
only one frequency. As a consequence, for a resonant cavity
($\overline \psi = 0$) and for a fixed optomechanical parameter
$\xi$, the standard quantum limit is reached at a single frequency
$\Omega_{\rm SQL}$ given by eq. (\ref{eq:zeta}),
\begin{equation}
M \Omega_{\rm SQL}^2 = \left|\chi\left[\Omega_{\rm
SQL}\right]\right|^{-1} = 2\hbar \xi^2. \label{eq:OmegaSQL}
\end{equation}
Curve {\it a} of Fig. \ref{fig:highfreq} shows the resulting noise
spectrum at resonance, which reaches the standard quantum limit
(dashed line) at frequency $\Omega_{\rm SQL}$. The radiation
pressure noise is dominant at lower frequency with a $1/\Omega^4$
dependence, whereas the constant phase noise limits the
sensitivity at higher frequency.

Curves {\it b} to {\it d} show the noise obtained for a detuned
cavity with the same optomechanical parameter $\xi$. Although these
curves exhibit a larger noise at low frequency than in the resonant
case, one gets a significant noise reduction below the standard
quantum limit in the intermediate frequency domain. The noise
reduction becomes larger and larger as the detuning increases. An
optimization of the noise spectrum (\ref{eq:Ssig}) for a given
detuning $\overline \psi$ and optomechanical parameter $\xi$ leads
to the optimal frequency $\Omega_{\rm min}$ and noise spectrum
$S_{\rm sig}^{\rm min}$,
\begin{eqnarray}
\Omega_{\rm min}^2 &=& \Omega_{\rm SQL}^2 \sqrt{1+\left(\overline
\psi/2\gamma\right)^2}, \label{eq:Omegahi} \\
\frac{S_{\rm sig}^{\rm min}\left[\Omega_{\rm min}\right]}{S_{\rm
sig}^{\rm SQL} \left[\Omega_{\rm min}\right]} &=&
\sqrt{1+\left(\overline \psi / 2\gamma\right)^2} - \overline \psi
/ 2\gamma. \label{eq:Ssighi}
\end{eqnarray}
The noise spectrum has an expression similar to eq.
(\ref{eq:Ssiglow2}) obtained at low frequency, except for the sign
of the detuning $\overline \psi$. As previously, the noise ratio
(\ref{eq:Ssighi}) tends to $\gamma/\overline \psi$ for large
detunings and one gets a noise reduction by a factor 10 for a
detuning $\overline \psi = 10\gamma$.

\section{Ultimate quantum limit} \label{sec:UQL}

The results of the previous sections seem to indicate that an
arbitrarily large sensitivity improvement can be obtained both in
the low and high frequency regimes since the equivalent input
noise evolves in both cases as $\gamma/\overline \psi$ for large
detunings. This actually is a consequence of the approximation
made on the mechanical susceptibility which was assumed to have no
imaginary part. It is possible to derive the optimal sensitivity
improvement at a given frequency $\Omega$ without any assumption
on the mechanical susceptibility $\chi \left[ \Omega \right]$. An
optimization of the noise spectrum (\ref{eq:Ssig}) with respect to
the optomechanical parameter $\xi$ leads to,
\begin{eqnarray}
\xi_{\rm min}^2 &=& \frac{\xi_{\rm SQL}^2\left[\Omega\right]}
{\sqrt{1+\left(\overline \psi/2\gamma\right)^2}},
\label{eq:xiUQL} \\
\frac{S_{\rm sig}^{\rm min}\left[\Omega\right]}{S_{\rm sig}^{\rm
SQL} \left[\Omega\right]} &=& \sqrt{1+\left(\overline \psi /
2\gamma\right)^2} + \frac{\overline \psi}{2\gamma} \frac{{\rm
Re}\left(\chi\left[\Omega\right]\right)}
{\left|\chi\left[\Omega\right]\right|}, \label{eq:SsigUQL}
\end{eqnarray}
where $\xi_{\rm SQL}\left[\Omega\right]$ is the optomechanical
parameter for which the standard quantum limit is reached at
frequency $\Omega$ for a resonant cavity [eq. (\ref{eq:xiSQL})].
As compared to eqs. (\ref{eq:xilow2}) and (\ref{eq:Ssiglow2})
obtained at low frequency and for a real mechanical
susceptibility, the only difference is the last term in eq.
(\ref{eq:SsigUQL}) which has a smaller amplitude when the
susceptibility has a non-zero imaginary part. As a consequence,
the equivalent input noise no longer decreases as
$\gamma/\overline \psi$ for very large detunings, and it reaches a
non-zero minimum value at a finite detuning given by,
\begin{eqnarray}
\overline \psi_{\rm min}/2\gamma &=& -\frac{{\rm
Re}\left(\chi\left[\Omega\right]\right)}{\left|{\rm Im}
\left(\chi\left[\Omega\right]\right)\right|}, \label{eq:psiUQL} \\
S_{\rm sig}^{\rm min}\left[\Omega\right] &=& \hbar \left|{\rm Im}
\left(\chi\left[\Omega\right]\right)\right|. \label{eq:SsigUQL2}
\end{eqnarray}
One then gets a limit to the sensitivity improvement which is only
related to the dissipation mechanism of the mechanical motion, via
the imaginary part of the susceptibility. This is nothing but the
ultimate quantum limit already predicted in the case of
interferometric measurements with squeezed-state injection
\cite{Jaekel90}. The same ultimate limit is thus reached by cavity
detuning.

\section{Cavity with a finite bandwidth} \label{sec:FiniteBandwidth}

\begin{figure}
\includegraphics[width=8cm]{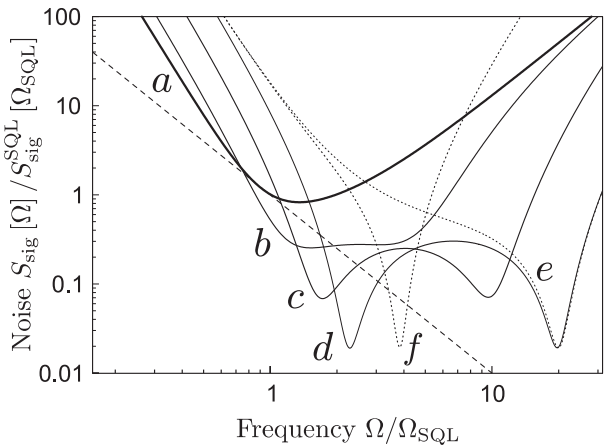}
\caption{Equivalent input noise $S_{\rm sig}\left[\Omega\right]$
at high frequency as a function of frequency $\Omega$, normalized
to the SQL values $\Omega_{\rm SQL}$ and $S_{\rm sig}^{\rm
SQL}\left[\Omega_{\rm SQL}\right]$. All curves are plotted for the
same optomechanical parameter $\xi$, and for the same finite
cavity bandwidth $\Omega_{\rm cav}=2\Omega_{\rm SQL}$, except
curve {\it f} for which $\Omega_{\rm cav}=\Omega_{\rm SQL}/3$.
Curves {\it a} to {\it d} correspond to positive normalized
detunings $\overline \psi/\gamma$ equal to 0, $2$, $5$, and $10$,
respectively. Curves {\it e} and {\it f} are obtained for a
negative normalized detuning of $-10$. The dashed line is the
standard quantum limit.} \label{fig:finitecav}
\end{figure}

We finally study the effect of a finite cavity bandwidth
$\Omega_{\rm cav}=\gamma/\tau$. The optical equations in the case
of a detuned cavity with a finite bandwidth are much more complex
than the ones given in the previous sections. As an example, the
input-output relation for the phase quadrature is derived from
eqs. (\ref{eq:InOut3}), (\ref{eq:quad_p}) and (\ref{eq:quad_q}),
\begin{eqnarray}
q^{\rm out}\left[\Omega\right] &=& 2\xi \frac{\gamma^2 +
\overline\psi^2 - i \gamma \Omega \tau}{\Delta} \left(X_{\rm
m}\left[\Omega\right]+X_{\rm sig}\left[\Omega\right]\right)
\nonumber \\
&&+\frac1{\Delta} \left(\gamma^2 + \overline\psi^2 + \Omega^2
\tau^2 \frac{\gamma^2 -\overline\psi^2}{\gamma^2
+\overline\psi^2}\right) q^{\rm in}\left[\Omega\right]
\nonumber \\
&&- \frac2{\Delta} \Omega^2 \tau^2 \frac{\gamma
\overline\psi}{\gamma^2 + \overline\psi^2} p^{\rm
in}\left[\Omega\right], \label{eq:qout3}
\end{eqnarray}
and can be compared to the simpler relation (\ref{eq:qout}) obtained
in the case of an infinite cavity bandwidth. We have computed the
equivalent input noise $S_{\rm sig}$ from the previous input-output
relation and from eqs. (\ref{eq:Frad_in}) to (\ref{eq:Frad_sig}), by
using the formal language Mathematica. Figure \ref{fig:finitecav}
shows the resulting noise obtained at high frequency, that is for a
mechanical susceptibility approximated by the real and negative
expression (\ref{eq:chihi}). All curves are plotted for the same
optomechanical parameter $\xi$ and, except for curve {\it f}, with a
cavity bandwidth $\Omega_{\rm cav}$ equal to $2 \Omega_{\rm SQL}$,
where $\Omega_{\rm SQL}$ is related to $\xi$ by eq.
(\ref{eq:OmegaSQL}).

Curve {\it a} of Fig. \ref{fig:finitecav} shows the equivalent input
noise at resonance ($\overline \psi = 0$). As compared to a cavity
with an infinite bandwidth (curve {\it a} of Fig.
\ref{fig:highfreq}), the noise is no longer constant at high
frequency but increases with the frequency. This is a consequence of
the low-pass filtering of the signal by the cavity for frequencies
larger than the cavity bandwidth. Curves {\it b} to {\it d} show the
equivalent noise spectrum obtained for positive and increasing
detunings. One clearly observes two resonant dips with a structure
very similar to the one already predicted for signal-recycled
gravitational-wave interferometers \cite{Buonanno01}. These two
resonances become deeper and more separated as the detuning
increases.

The dip at the lowest frequency is very similar to the one
obtained for an infinite cavity bandwidth, as well in frequency
position, in width, and in noise reduction (compare curves {\it b}
to {\it d} of Figs. \ref{fig:highfreq} and \ref{fig:finitecav}).
In both cases, the dip can be associated with the resonance of the
amplification factor $\left|\chi_{\rm eff}/\chi\right|$. From eq.
(\ref{eq:chieff}), one indeed finds that the effective
susceptibility $\chi_{\rm eff}$ has a Lorentzian shape with a
resonance frequency $\Omega_-$ very close to the dip position
$\Omega_{\rm min}$ [eq. (\ref{eq:Omegahi})] and given for a large
detuning by,
\begin{equation}
\Omega_- \simeq \Omega_{\rm SQL} \sqrt{\frac{\overline
\psi}{2\gamma}}. \label{eq:OmegaEff}
\end{equation}
Taking a finite cavity bandwidth thus changes the width of the
effective mechanical resonance, as already discussed in section
\ref{sec:Mirror} [see eq. (\ref{eq:Gammaeff})], but it has no
apparent effect on the sensitivity improvement around the
resonance frequency $\Omega_-$.

The second dip only exists for a finite cavity bandwidth and is a
consequence of the optical dynamics in the cavity. Its frequency
actually corresponds to the resonance frequency $\Omega_+$ of the
term $1/\Delta$ which appears both in the input-output relation
(\ref{eq:qout3}) and in the radiation pressure forces
(\ref{eq:Frad_in}) to (\ref{eq:Frad_sig}),
\begin{equation}
\Omega_+ = \Omega_{\rm cav} \sqrt{1+\frac{\overline
\psi^2}{\gamma^2}}. \label{eq:OmegaDelta}
\end{equation}
In contrast to the first dip for which the signal amplification is
only obtained with a positive detuning, the second dip exists both
for positive and negative detunings. This is clearly visible in
Fig. \ref{fig:finitecav} where curves {\it d} and {\it e} are
plotted for the same parameters but for reverse detunings. Note
however that the stability conditions are very different in the
two situations. In particular the dynamic stability condition
(\ref{eq:Stability2}) is always satisfied for a negative detuning
whereas it is very restrictive for a positive detuning. Curves
{\it b} to {\it d} of Fig. \ref{fig:finitecav} are actually
unstable for a reasonably not too large mechanical damping
$\Gamma$.

The sensitivity improvement at the resonance frequencies
$\Omega_\pm$ can be computed from the analytic expression given by
Mathematica. One gets for a large detuning $\overline \psi$,
\begin{equation}
\frac{S_{\rm sig}\left[\Omega_\pm\right]}{S_{\rm sig}^{\rm
SQL}\left[\Omega_{\rm SQL}\right]} \simeq \frac{2\gamma^2}
{\overline \psi^2}. \label{eq:Ssigcav}
\end{equation}
The two dips have thus the same depth, as it can be observed in
Fig. \ref{fig:finitecav}. In this expression, the noise is
normalized to the standard quantum limit at frequency $\Omega_{\rm
SQL}$. It is also of interest to compute the noise reduction below
the standard quantum limit, that is the ratio between the noise at
frequency $\Omega_\pm$ and the standard quantum limit
(\ref{eq:SsigSQL}) at the same frequency,
\begin{equation}
\frac{S_{\rm sig}\left[\Omega_\pm\right]}{S_{\rm sig}^{\rm
SQL}\left[\Omega_\pm\right]} \simeq \frac{2 \gamma^2}{\overline
\psi^2} \left(\frac{\Omega_\pm}{\Omega_{\rm SQL}}\right)^2.
\label{eq:Ssigcav2}
\end{equation}
From eq. (\ref{eq:OmegaEff}), the ratio tends to $\gamma/\overline
\psi$ at the resonance frequency $\Omega_-$ of the first dip. This
result is identical to the one obtained in section
\ref{sec:HighFreq} for an infinite cavity bandwidth. At the
resonance frequency $\Omega_+$ of the second dip [eq.
(\ref{eq:OmegaDelta})], the noise ratio (\ref{eq:Ssigcav2}) is
equal to $2 \left(\Omega_{\rm cav}/\Omega_{\rm SQL}\right)^2$. The
noise is then reduced below the standard quantum limit only if the
cavity bandwidth is small enough, as shown by curves {\it e} and
{\it f} in Fig. \ref{fig:finitecav}, respectively obtained for
$\Omega_{\rm cav} = 2\Omega_{\rm SQL}$ and $\Omega_{\rm cav} =
\Omega_{\rm SQL}/3$.

Finally note that eq. (\ref{eq:Ssigcav}) seems to indicate that an
arbitrarily small equivalent input noise can be reached by
increasing the detuning. As for an infinite cavity bandwidth
(section \ref{sec:UQL}), it can be shown that the noise is always
larger than the ultimate quantum limit (\ref{eq:SsigUQL2}), which
can be reached at every frequency $\Omega$ by an appropriate
choice of the parameters $\xi$, $\overline \psi$, and $\Omega_{\rm
cav}$.

\section{Conclusion}

We have studied the quantum limits of an optomechanical sensor based
on a detuned high-finesse cavity with a movable mirror. We have
shown that the sensitivity to a variation of the cavity length can
be improved beyond the standard quantum limit, up to the ultimate
quantum limit which only depends on the dissipation mechanisms of
the mirror motion. This improvement is due to an amplification of
the signal by the mirror displacements. The coupling between the
mirror motion and the intracavity light field actually changes the
dynamics of the mirror, both via its spring constant and its
damping. But the mirror motion also becomes sensitive to the signal
and can amplify the effect of the signal on the intracavity field.
For a finite cavity bandwidth, one gets a sensitivity improvement
very similar to the one predicted in signal-recycled
gravitational-waves interferometers, with two dips in the equivalent
input noise which are related to the effective mechanical resonance
of the mirror and to the optical dynamics in the cavity. A
high-finesse cavity with a movable mirror thus appears as a model
system to test quantum effects in large-scale interferometers.

\acknowledgments

We thank Jean-Michel Courty and Julien Le Bars for fruitful
discussions. This work was partially funded by EGO (collaboration
convention EGO-DIR-150/2003 for a study of quantum noises in
gravitational waves interferometers).

\end{document}